\documentclass[journal=jpcbfk,layout=twocolumn,manuscript=article]{achemso}

\usepackage[utf8]{inputenc}
\usepackage[T1]{fontenc}
\usepackage[version=3]{mhchem}
\usepackage[dvipsnames,x11names]{xcolor}
\usepackage{graphicx}
\usepackage{soul}
\usepackage{times}
\usepackage{calrsfs}
\usepackage[colorlinks=true,plainpages=false]{hyperref}
\hypersetup{pdftitle={Editorial}, 
            colorlinks=true,
            citecolor = {cyan},
	        linkcolor = {blue},
            urlcolor  = {purple}
}
\usepackage{orcidlink}
\usepackage[normalem]{ulem}
\usepackage{mathtools}
\usepackage{amsmath}
\usepackage{amssymb}
\usepackage{threeparttable}
\mciteErrorOnUnknownfalse

\newcommand*{\figref}[1]{%
  \hyperref[{#1}]{%
    Figure~\ref*{#1}%
  }%
}
\renewcommand*{\eqref}[1]{%
  \hyperref[{#1}]{%
    eq~\ref*{#1}%
  }%
}
\newcommand*{\tabref}[1]{%
  \hyperref[{#1}]{%
    Table~\ref*{#1}%
  }%
}
\newsavebox{\verbbox}
\usepackage{caption}
\usepackage{float}
\usepackage{stfloats}

\author{Christopher Kang \orcidlink{0000-0002-9245-8101}}
\affiliation[NC]
{{Department of Chemical and Biomolecular Engineering, North Carolina State University, Raleigh, North Carolina 27695, USA}}

\author{Rahul Verma \orcidlink{0000-0003-4991-8376}}
\affiliation[NC]
{{Department of Chemical and Biomolecular Engineering, North Carolina State University, Raleigh, North Carolina 27695, USA}}

\author{Aditya Sonpal \orcidlink{0000-0003-4317-1654}}
\affiliation[NC]
{{Department of Chemical and Biomolecular Engineering, North Carolina State University, Raleigh, North Carolina 27695, USA}}

\author{Alyson Shoji \orcidlink{XXXX-XXXX-XXXX-XXXX}}
\affiliation[NC]
{{Department of Chemical and Biomolecular Engineering, North Carolina State University, Raleigh, North Carolina 27695, USA}}

\author{Christophe Chipot \orcidlink{0000-0002-9122-1698}}
\affiliation[CNRS]{Laboratoire de Physique et Chimie Théoriques, 
                   Unité Mixte de Recherche n\textsuperscript{o}7019, 
                   Université de Lorraine, 
                   54506 Vand\oe uvre-l\`es-Nancy Cedex, France}
\alsoaffiliation[TCBG]{Theoretical and Computational Biophysics Group, Beckman Institute,  
                       University of Illinois at Urbana-Champaign, 
                       Urbana, Illinois 61801, USA}
\alsoaffiliation[UIUC]{Department of Physics, 
                       University of Illinois at Urbana-Champaign, 
                       Urbana, Illinois 61801, USA}
\alsoaffiliation[UC]{Department of Biochemistry and Molecular Biology, 
                     The University of Chicago, 
                     Chicago, Illinois 60637, USA}
\alsoaffiliation[UH]{Department of Chemistry, 
                     The University of Hawai`i at M\={a}noa, Honolulu, Hawaii 96822, USA}

\author{Jim Pfaendtner \orcidlink{0000-0001-6727-2957}}
\affiliation[NC]
{{Department of Chemical and Biomolecular Engineering, North Carolina State University, Raleigh, North Carolina 27695, USA}}

\email{wjpfaendtner@ncsu.edu}

\title{A Force-Kernel Reformulation of the Extended-System Adaptive Biasing Force for Free-Energy Calculations}

\begin{document}

\maketitle

\clearpage
\section{Abstract}
We introduce force-kernel extended-system adaptive biasing force (FK-eABF), a force-based kernel reformulation of eABF that replaces the histogram-based mean-force accumulator of conventional eABF with a sparse population of Gaussian kernels storing local running-mean forces. Biasing forces are recovered by Nadaraya--Watson regression, yielding smooth estimates from the earliest stages of a simulation without a minimum-count threshold, while the same kernel population also defines an auxiliary, self-attenuating exploration force that requires no prior knowledge of barrier heights. On $N$-acetyl-$N^\prime$-methylalanylamide in explicit water, FK-eABF achieves full free-energy landscape coverage faster than well-tempered metadynamics (WT-MetaD), on-the-fly probability enhanced sampling (OPES), and WTM-eABF, while all four methods converge to comparable accuracy given sufficient time. FK-eABF also retains long-time accuracy: on the DFG-in/out transition of Abl1 kinase, multi-microsecond simulations recover the established near-isoenergetic balance between states. At the opposite extreme, applied to the electrocyclic ring closure of 1,3-butadiene at the ab initio molecular dynamics level, FK-eABF recovers the free-energy landscape within 30 ps. Together, these benchmarks, spanning more than four orders of magnitude in simulation time, establish FK-eABF as more than a kernelized implementation of eABF: A force-based kernel reformulation that delivers faster early-time convergence without sacrificing long-time quantitative accuracy.

\section{Introduction}

Free-energy calculations along collective variables (CVs) are a cornerstone of computational biophysics, providing quantitative insight into the thermodynamics of ligand binding,\cite{gumbart2013standard,fu2021} conformational change,\cite{mora2013} membrane permeation,\cite{tse2018link,kang2024impact} and other rare events that are inaccessible to conventional, unbiased molecular dynamics (MD) simulations.\cite{zimmerman2021sars,lindorff2011fast} The most widely employed CV-based free-energy methods---umbrella sampling,\cite{Torrie1977a} metadynamics,\cite{Laio2002} the adaptive biasing force (ABF),\cite{Darve2001} and their many respective variants---each embody a distinct philosophy for driving exploration of CV space, and each carry different practical tradeoffs in how quickly a reliable free-energy estimate becomes available, and how flexibly the underlying representation adapts as sampling accumulates.

Among these, the extended-system ABF (eABF) formulation\cite{lesage2017,fu2016} has emerged as a particularly flexible framework, precisely because it separates the biasing mechanism from the free-energy estimator. In eABF, a fictitious variable $\boldsymbol{\lambda}$ is coupled harmonically to $\mathbf{z}=\boldsymbol{\xi}(\mathbf{q})$, and the bias acts exclusively on $\boldsymbol{\lambda}$, thereby eliminating the need for Jacobian and second-derivative estimates of the CV, and permitting the free energy to be recovered by the corrected $z$-averaged restraint (CZAR) estimator,\cite{lesage2017} which remains asymptotically exact, irrespective of the applied bias. The robustness of this architecture has motivated a growing family of hybrid methods, e.g., well-tempered metadynamics-eABF (WTM-eABF),\cite{fu2019} and on-the-fly probability enhanced sampling (OPES)-eABF,\cite{hulm2025combining} among other variants,\cite{zhou2025one} that combine the eABF force accumulator and CZAR estimator with different exploration strategies, treating the force representation and the exploration drive as modular components.\cite{fu2019,fu2020}

Despite this modularity, limitations remain in how the mean force is represented and how exploration is driven. In conventional eABF implementations, mean-force estimates are accumulated on a discretized grid, leading to the well-known tradeoff between spatial resolution and statistical efficiency; in higher dimensions, the exponential growth of the grid elements renders sparse sampling a practical bottleneck. The metadynamics component of WTM-eABF accelerates barrier crossing, but introduces nontrivial parameter sensitivity in the hill height, width, and bias temperature that must be balanced carefully.\cite{barducci2008,bussi2020using} Moreover, because individual Gaussian hills are never revised once deposited, errors incurred during early, poorly converged sampling are irreversibly encoded in the bias, even as the deposition rate attenuates.\cite{dama2014well} 

The OPES method\cite{invernizzi_rethinking_2020} addresses these limitations through a fundamentally different paradigm. Rather than summing permanent deposits, it reconstructs the bias at each step from a reweighted kernel-density estimate of the unbiased distribution, reaching a quasi-static regime in which the bias shape stabilizes and early errors are progressively diluted. This architecture often yields faster and more robust convergence than hill-based deposition, but introduces a dependence on a barrier parameter that sets the target distribution, and like all kernel-density methods, its accuracy in sparsely sampled regions remains contingent on the fidelity of the density estimate.

In the present work, we introduce force-kernel eABF (FK-eABF), which retains the extended-system architecture and CZAR estimator of eABF, but reformulates the representation problem at the level of force estimation rather than probability reconstruction. The histogram accumulator is replaced with a compressed population of Gaussian kernels. The compression framework is shared with OPES,\cite{invernizzi_rethinking_2020} but the two methods differ in what is stored per kernel. OPES kernels carry reweighted heights that sum into a probability estimate and measure inter-kernel distances using each candidate kernel's own bandwidth, whereas FK-eABF kernels carry running-mean force estimates that enter a Nadaraya--Watson (NW) regression\cite{nadaraya1964estimating,watson1964smooth}. The resulting bias force is a smooth, continuous gradient field recovered as a locally weighted average of force measurements, that can represent arbitrary shapes with a finite number of kernels. Since each kernel's influence extends over a region set by its bandwidth, the force estimate at any point borrows statistical strength from neighboring kernels, providing a usable mean-force estimate in sparsely sampled regions where a histogram bin would still be empty. As sampling accumulates and the adaptive kernel bandwidth contracts, the representation progressively resolves finer spatial features, so the early-time advantage does not come at the cost of late-time resolution.

To accelerate early exploration, FK-eABF optionally  introduces a density-based force derived from the same $\lambda$-kernel population that supplies the ABF cancellation force. This auxiliary drive repels $\boldsymbol{\lambda}$ from already well-sampled regions, acting as a self-limiting exploration mechanism that is strongest during the early stages of sampling and naturally attenuates as CV-space coverage becomes uniform. Since it is reconstructed on-the-fly from the evolving kernel population, rather than accumulated through permanent deposits, early errors in the force estimate are progressively corrected as kernel statistics converge. At the same time, because the exploration force acts only on $\boldsymbol{\lambda}$, it does not enter the CZAR estimator directly, thereby enhancing sampling efficiency without biasing the recovered free energy estimate.

\begin{figure}[!ht]
\centering
\includegraphics[width=\columnwidth]{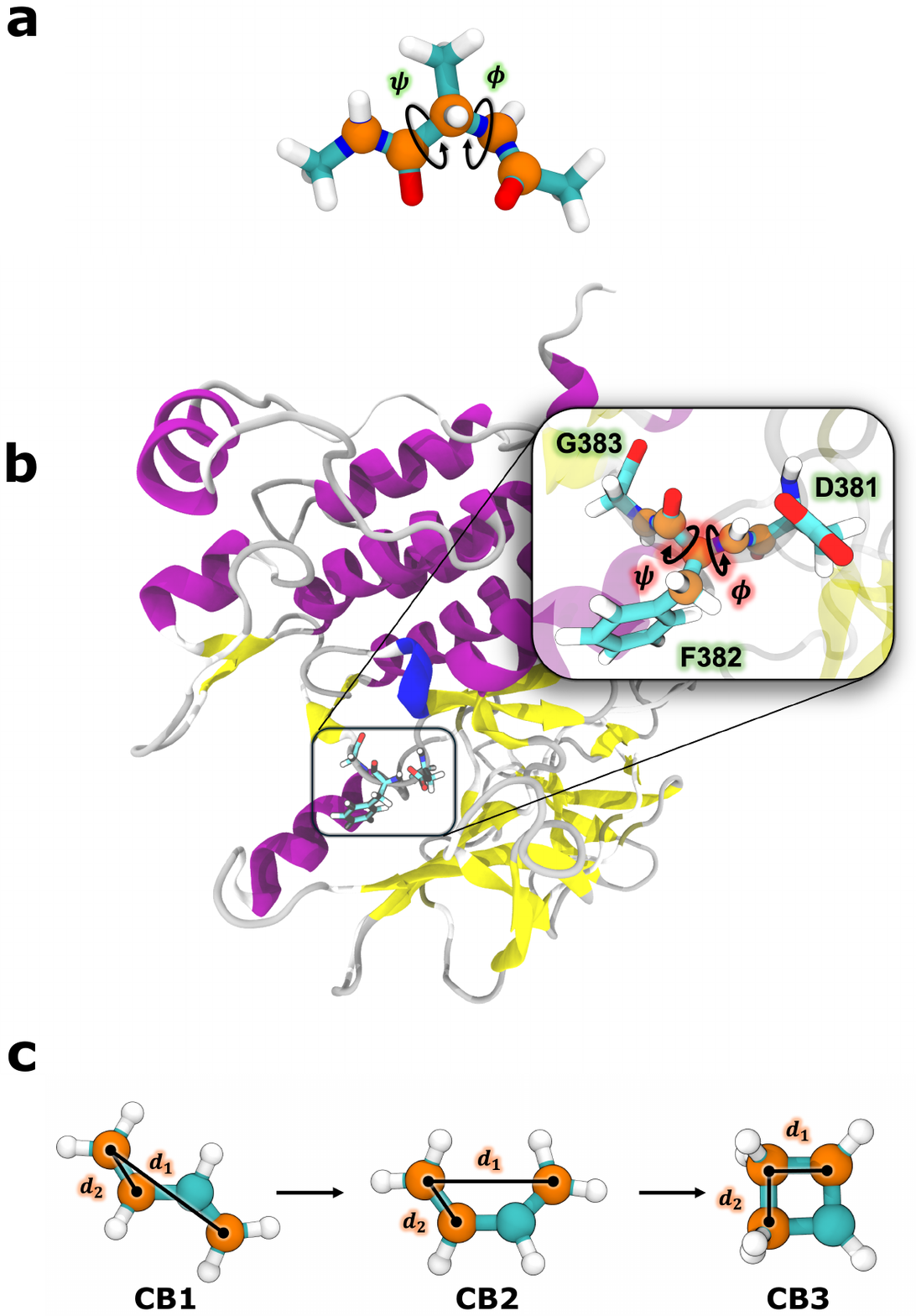}
\caption{Molecular systems and collective variables employed in this work. (a) $N$-acetyl-$N^\prime$-methylalanylamide in explicit water. (b) Apo BCR-Abl1 kinase with the DFG motif (Asp381--Phe382--Gly383) highlighted. (c) Electrocyclic ring closure of \textit{trans}-1,3-butadiene (CB1) through \textit{cis}-1,3-butadiene (CB2) to cyclobutene (CB3). CV atoms are shown as orange spheres throughout.}
\label{fig:systems}
\end{figure}

We validate FK-eABF on two benchmark systems---the two-dimensional M\"uller--Brown potential, where an analytical reference enables exact convergence assessment, and $N$-acetyl-$N^\prime$-methylalanylamide (NANMA), also known as alanine dipeptide, in explicit water (\figref{fig:systems}a), where we benchmark against well-tempered metadynamics (WT-MetaD), OPES, and WTM-eABF under identical simulation conditions---and further evaluate its performance on two systems that probe distinct practical challenges, namely the DFG-in/out conformational transition of Abl1 kinase,\cite{oruganti_allosteric_2022, shan_conserved_2009,meng2015computational} a biologically relevant process with a rugged high-dimensional free-energy landscape and slow orthogonal relaxation (\figref{fig:systems}b), and the ring closure of 1,3-butadiene at the \emph{ab initio} MD level,\cite{awasthi2017exploring} where the high computational cost per time step strongly favors methods with rapid early-time convergence (\figref{fig:systems}c).

\section{Theoretical Basis}

\paragraph{\textbf{Extended-system adaptive biasing force.}}
FK-eABF builds on the eABF framework of Lesage et al.\cite{lesage2017} In eABF, a $d$-dimensional fictitious variable $\boldsymbol{\lambda} \in \mathbb{R}^d$ is harmonically coupled to the CV, $\mathbf{z} = \boldsymbol{\xi}(\mathbf{q})$ through the extended potential
\begin{equation}
V_\mathrm{ext}(\mathbf{q},\boldsymbol{\lambda}) = V(\mathbf{q}) + \sum_{i=1}^{d}\frac{\kappa_i}{2}\bigl(\xi_i(\mathbf{q}) - \lambda_i\bigr)^2 ,
\label{eq:Vext}
\end{equation}
where $V(\mathbf{q})$ is the physical potential energy and $\kappa_i$ is the spring constant along dimension~$i$. The fictitious particle evolves under Langevin dynamics at the system temperature~$T$, with $\beta = (k_{\rm B}T)^{-1}$.

The running estimate of the mean spring force $\langle \kappa_i(z_i-\lambda_i)\rangle_{\boldsymbol{\lambda}}$ is used as a biasing force acting on $\boldsymbol{\lambda}$, thereby progressively flattening the extended free energy $A_\kappa(\boldsymbol{\lambda})$. Since the bias acts only on $\boldsymbol{\lambda}$, the physical coordinates are not directly subjected to the biasing force. The physical free energy $A(\mathbf{z})$ is then recovered via the CZAR estimator,\cite{lesage2017}
\begin{equation}
\frac{\partial A}{\partial z_i}
= -\langle\kappa_i(z_i - \lambda_i)\rangle_{\mathbf{z}}
  - \beta^{-1}\,\frac{\partial}{\partial z_i}
    \ln\tilde\rho(\mathbf{z}),
\label{eq:czar}
\end{equation}
where $\tilde\rho(\mathbf{z})$ is the biased marginal distribution and $\langle\cdot\rangle_{\mathbf{z}}$ denotes an average conditioned on $\mathbf{z}$.  Integrating eq~\ref{eq:czar} on a grid yields $A(\mathbf{z})$. In conventional eABF, both terms are accumulated in fixed-width histograms, but in FK-eABF, both are recovered from a compressed kernel representation.

\paragraph{\textbf{Force-kernel compression.}}
FK-eABF replaces the histogram-based accumulators of conventional eABF with a compressed representation based on Gaussian kernels, using a compression strategy analogous to OPES,\cite{invernizzi_rethinking_2020} but applied to  local force estimates rather than bias-potential contributions. Two independent kernel populations are maintained: $\lambda$-kernels in the fictitious-variable space, which define the bias, and $z$-kernels in the real CV space, which enter the CZAR estimator.

Each kernel $k$ stores a center $\mathbf{c}_k$, a kernel-averaged force $\boldsymbol{\mu}_k$, a bandwidth $\boldsymbol{\sigma}_k$, and a sample count $N_k$. The Gaussian window function is
\begin{equation}
G_k(\mathbf{s}) = \exp\!\left(-\sum_{i=1}^{d}\frac{(s_i - c_{k,i})^2}{4\,\sigma_{k,i}^2}\right) ,
\label{eq:gaussian}
\end{equation}
where $\mathbf{s}$ is an evaluation point ($\boldsymbol{\lambda}$ or $\mathbf{z}$). With the convention of eq. \ref{eq:gaussian}, the effective Gaussian standard deviation is $\sqrt{2}\sigma_{k,i}$. We adopt this form because the broader window enhances the smoothness of the reconstructed force and density fields during early sampling. As $\sigma_{k,i}$ varies across kernels, each kernel's contribution is weighted by a normalization factor
\begin{equation}
\alpha_k = \prod_{j=1}^{d} \frac{\sigma_{0,j}}{\sigma_{k,j}} \,,
\label{eq:alpha}
\end{equation}
where $\sigma_{0,i}$ is the initial bandwidth. This factor restores the relative normalization of variable-bandwidth kernels, so that narrow versus broad kernels enter the reconstructed density with the appropriate relative weights, up to an overall constant factor. Additional details of $\alpha_k$ are provided in the Supporting Information.

As data accumulates, the global bandwidth $\sigma_{\mathrm{global},i}$ contracts according to Silverman's rule,\cite{silverman2018density}
\begin{equation}
\sigma_{\mathrm{global},i} = \sigma_{0,i}\!\left(\frac{n_\mathrm{eff}(d+2)}{4}\right)^{\!-1/(4+d)} ,
\label{eq:silverman}
\end{equation}
where $n_\mathrm{eff} = (\sum_k N_k)^2/\sum_k N_k^2$ is the effective sample size, which reduces to the number of kernels $M$ when all carry equal counts. A user-specified floor $\sigma_\mathrm{min}$ ensures that the bandwidth does not contract below a physically meaningful resolution.

When a new force sample ($f_i = \kappa_i(z_i - \lambda_i)$) arrives, the algorithm identifies the nearest kernel according to the normalized distance
\begin{equation}
d^2(\mathbf{s}, k) = \sum_{i=1}^{d}\frac{(s_i - c_{k,i})^2}{4\,\sigma_{\mathrm{global},i}^2}.
\label{eq:merge_dist}
\end{equation}
This merge distance is defined using the global bandwidth $\sigma_{\mathrm{global},i}$ rather than the per-kernel width---in deliberate contrast to OPES, which uses the per-kernel width of the candidate. Using a global metric avoids a feedback loop, whereby contracting widths inhibit merging and artificially inflate the kernel count. If $d^2 < \theta^2/4$ (with compression threshold $\theta = 1.0$ in all simulations reported here), the sample is absorbed, otherwise, a new kernel is instantiated.

Upon absorption, the center and mean force are updated as weighted means,
\begin{equation}
\left\{
\begin{array}{lll}
c_{k,i} & \leftarrow & \displaystyle
\frac{N_{\mathrm{old}}\,c_{k,i} + s_i}{N_{\mathrm{old}} + 1},
\\[0.5cm]
\mu_{k,i} & \leftarrow & \displaystyle
\frac{N_\mathrm{old}\,\mu_{k,i} + f_i}{N_\mathrm{old} + 1} ,
\label{eq:updates}
\end{array}
\right.
\end{equation}
and the bandwidth $\sigma_{k,i}$ is updated using the pairwise variance formula of Chan et al.\cite{chan1983algorithms} Each incoming sample is treated as a Gaussian of width $\sigma_{\mathrm{global},i}$, rather than as a point, so $\sigma_{k,i}$ represents an effective local bandwidth, namely, the combined width of all absorbed Gaussians, rather than a literal sample variance. This ensures that even a single-sample kernel has a well-defined spatial influence radius. After absorption, the algorithm recursively searches for additional merge candidates until none remain. Note that the current implementation supports up to three CVs, a limit set by the grid-based interpolation of the mean-force field rather than the kernel representation itself; the algorithm is general for arbitrary dimensionality $d$.

\paragraph{\textbf{Bias force from $\lambda$-kernels.}}
At any point $\boldsymbol{\lambda}$, the mean force is evaluated using Nadaraya--Watson (NW) kernel regression:\cite{nadaraya1964estimating,watson1964smooth}
\begin{equation}
\hat{g}_i(\boldsymbol{\lambda}) = \frac{\displaystyle\sum_{k=1}^{M} \alpha_k\,N_k\,G_k(\boldsymbol{\lambda})\,\mu_{k,i}}{\displaystyle\sum_{k=1}^{M} \alpha_k\,N_k\,G_k(\boldsymbol{\lambda})} .
\label{eq:nw}
\end{equation}
Only the relative variation of $\alpha_k$ across kernels affects $\hat{g}_i(\boldsymbol{\lambda})$: a global multiplicative rescaling cancels between numerator and denominator of the NW ratio, whereas kernel-dependent variation arising from differing $\sigma_k$ reweights narrow versus broad kernels in the average. The ABF cancellation force is then $F_{\mathrm{ABF},i} = -\hat{g}_i(\boldsymbol{\lambda})$.

Unlike a histogram, which confines information to fixed bins, the NW estimator leverages contributions from neighboring kernels, providing smooth force estimates even in sparsely sampled regions. Unlike OPES, which constructs a biasing potential by summing Gaussian hills into a bias potential, eq~\ref{eq:nw} directly yields a continuous gradient field via local averaging.

\paragraph{\textbf{Density-based exploration.}}
Before convergence, $\boldsymbol{\lambda}$ may remain trapped near its initial basin. FK-eABF optionally adds a repulsive exploration force derived from the NW denominator,
\begin{equation}
Z(\boldsymbol{\lambda}) = \sum_{k=1}^{M} \alpha_k\,N_k\, G_k(\boldsymbol{\lambda}),
\label{eq:Z}
\end{equation}
which quantifies the local kernel support. The exploration potential is,
\begin{equation}
\left\{
\begin{array}{rll}
V_\mathrm{ex}(\boldsymbol{\lambda}) & = & \displaystyle c\,\ln\!\left(1 + \frac{Z(\boldsymbol{\lambda})}{Z_0}\right),
\\[0.5cm]
c & = & \beta^{-1} (\gamma - 1),
\label{eq:Vex}
\end{array}
\right.
\end{equation}
which generates the force,
\begin{equation}
F_{\mathrm{ex},i}
= -\frac{c}{Z_0 + Z(\boldsymbol{\lambda})}\,
\frac{\partial Z}{\partial\lambda_i},
\label{eq:Fex}
\end{equation}
where $\gamma \ge 1$ is a bias factor and $Z_0$ is the median of $Z$ over populated grid nodes, with a floor of $Z_0 = 1$ at early times. The median provides a robust estimate of typical kernel support, avoiding domination by the most heavily sampled basin. The floor regularizes the earliest stage of sampling before reliable statistics accumulate, and $Z_0$ is updated only on grid rebuilds to prevent short-time fluctuations from modulating the exploration drive.

The exploration force has three key properties. First, it is \emph{self-limiting}: for $Z \gg Z_0$, $Z_0$ becomes negligible and the force saturates as $F_{\mathrm{ex},i} \approx -c\,\partial_i\ln Z$, making the exploration drive insensitive to the precise value of $Z_0$ wherever sampling has already occurred. Second, it is \emph{self-attenuating}: As the ABF cancellation converges and the sampling distribution approaches uniformity, $\nabla Z$ diminishes globally and the exploration force decays toward zero. Third, it is \emph{reconstructive}: $Z$ and its gradient are recomputed from the instantaneous kernel population at each grid update, so that errors from early sampling are diluted as kernel means converge, rather than being irreversibly embedded in a deposited bias. The exploration force acts only on $\boldsymbol{\lambda}$ and does not enter the CZAR estimator directly.

\paragraph{\textbf{Free-energy recovery from $z$-kernels.}}
The $z$-kernel population stores the unclamped spring force $\kappa_i(z_i - \lambda_i)$ at locations in the real CV space. Applying the same NW estimator to this population yields both contributions to the CZAR identity (eq~\ref{eq:czar}):
\begin{equation}
\begin{split}
\frac{\partial A}{\partial z_i} =
&-\frac{\sum_k \alpha_k\, N_k\, G_k(\mathbf{z})\,\mu_{k,i}}
       {\sum_k \alpha_k\, N_k\, G_k(\mathbf{z})}\\
\;&-\; \beta^{-1}\,\frac{\partial}{\partial z_i}
\ln\!\left(\sum_k \alpha_k\, N_k\, G_k(\mathbf{z})\right),
\label{eq:czar_kernel}
\end{split}
\end{equation}
where the sums run over the $M_z$ $z$-kernels. The role of $\alpha_k$ is the same as in eq~\ref{eq:nw}. Only its relative variation across kernels matters. A global multiplicative rescaling leaves both the normalized force average and the density-gradient term unchanged, whereas kernel-dependent variation in $\alpha_k$ reweights narrow versus broad kernels in both contributions. Thus, the same compressed $z$-kernel population supplies both the conditional spring-force estimate and the density-gradient term, after which $A(\mathbf{z})$ is obtained by numerically integrating eq~\ref{eq:czar_kernel} on a grid. In the limit of exhaustive sampling and vanishing bandwidth, with compression and grid-integration errors also vanishing, the kernel estimator approaches the exact conditional mean force and the kernel CZAR expression recovers the exact free-energy gradient. Under these limiting conditions, the asymptotic exactness of ABF is preserved in the kernel representation.

\begin{figure*}[!ht]
\centering
\includegraphics[width=\textwidth]{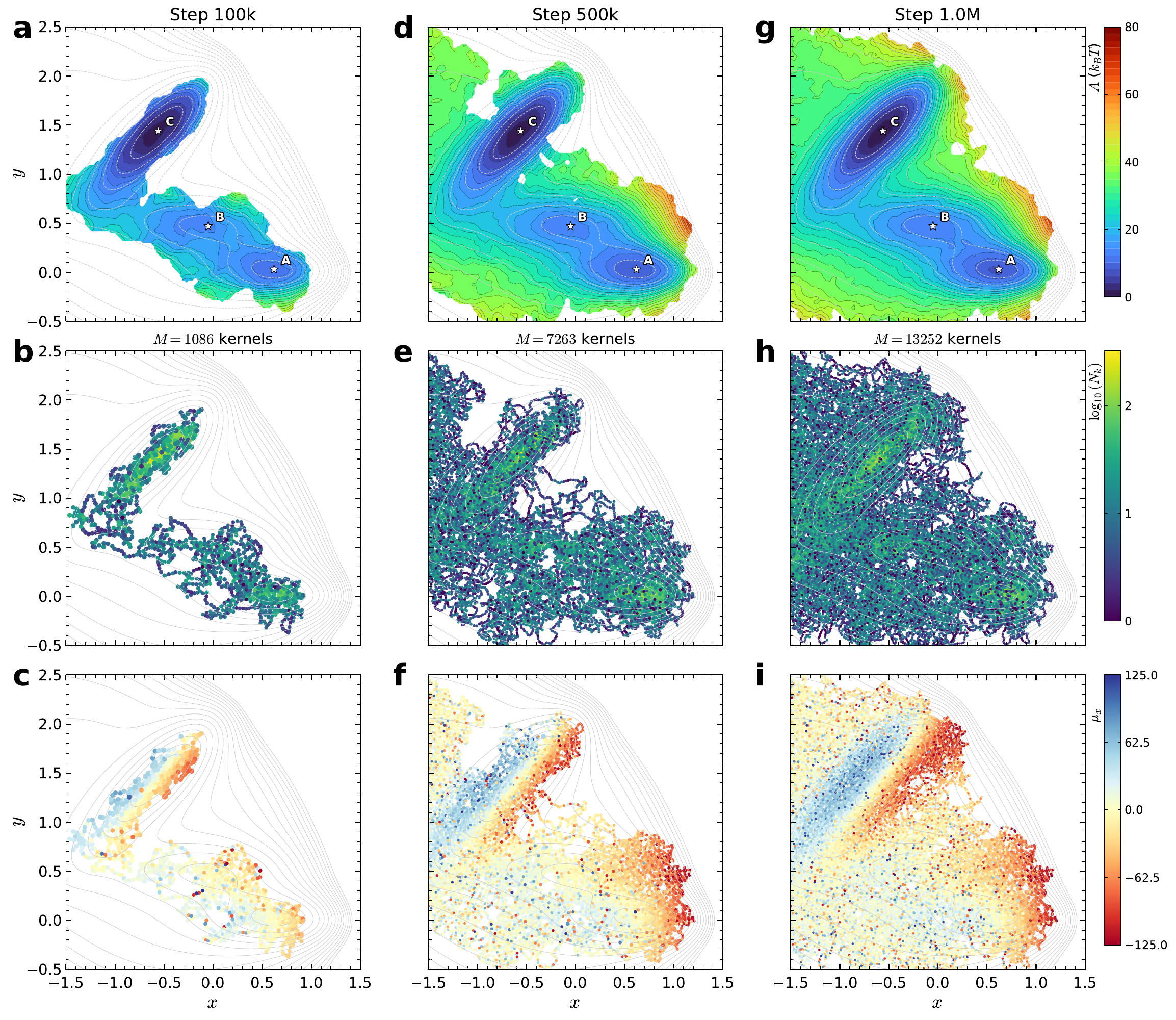}
\caption{Time evolution of the force-kernel population and recovered free energy on the M\"uller--Brown potential. Columns correspond to three simulation time points.  \textbf{Top row:} CZAR-recovered landscape with analytical contour lines.  \textbf{Middle row:} $\lambda$-kernel ellipses colored by $\log_{10}(N_k)$. \textbf{Bottom row:} Same kernel ellipses colored by the signed $x$-component of the stored mean force, $\mu_{k,x}$.}
\label{fig:kernel_evolution}
\end{figure*}

\section{Results}

\paragraph{\textbf{M\"uller}--Brown potential.} The two-dimensional M\"uller--Brown potential, scaled by a prefactor of 0.2, was used to assess the accuracy and convergence properties of FK-eABF. This surface features three distinct minima (A, B and C; \figref{fig:kernel_evolution}) separated by barriers on the order of 5--15 $k_{\rm B}T$, providing a stringent test of the ability of the method to resolve competing basins, transition regions and barrier heights from a single continuous trajectory. The simulation was performed using potential energy surface MD as implemented in PLUMED\cite{tribello2014plumed} for a total of $10^7$ steps. The M\"uller--Brown simulation was carried out without density-based exploration ($\gamma=1$), thereby isolating the convergence properties of the base method. In addition to validating the recovered free-energy surface, this benchmark also verifies that the biased $z$-kernel density estimate entering the second term of the kernel CZAR identity remains well behaved during early and intermediate stages of sampling (Figure~S1).

\begin{figure*}[!hb]
\centering
\includegraphics[width=\textwidth]{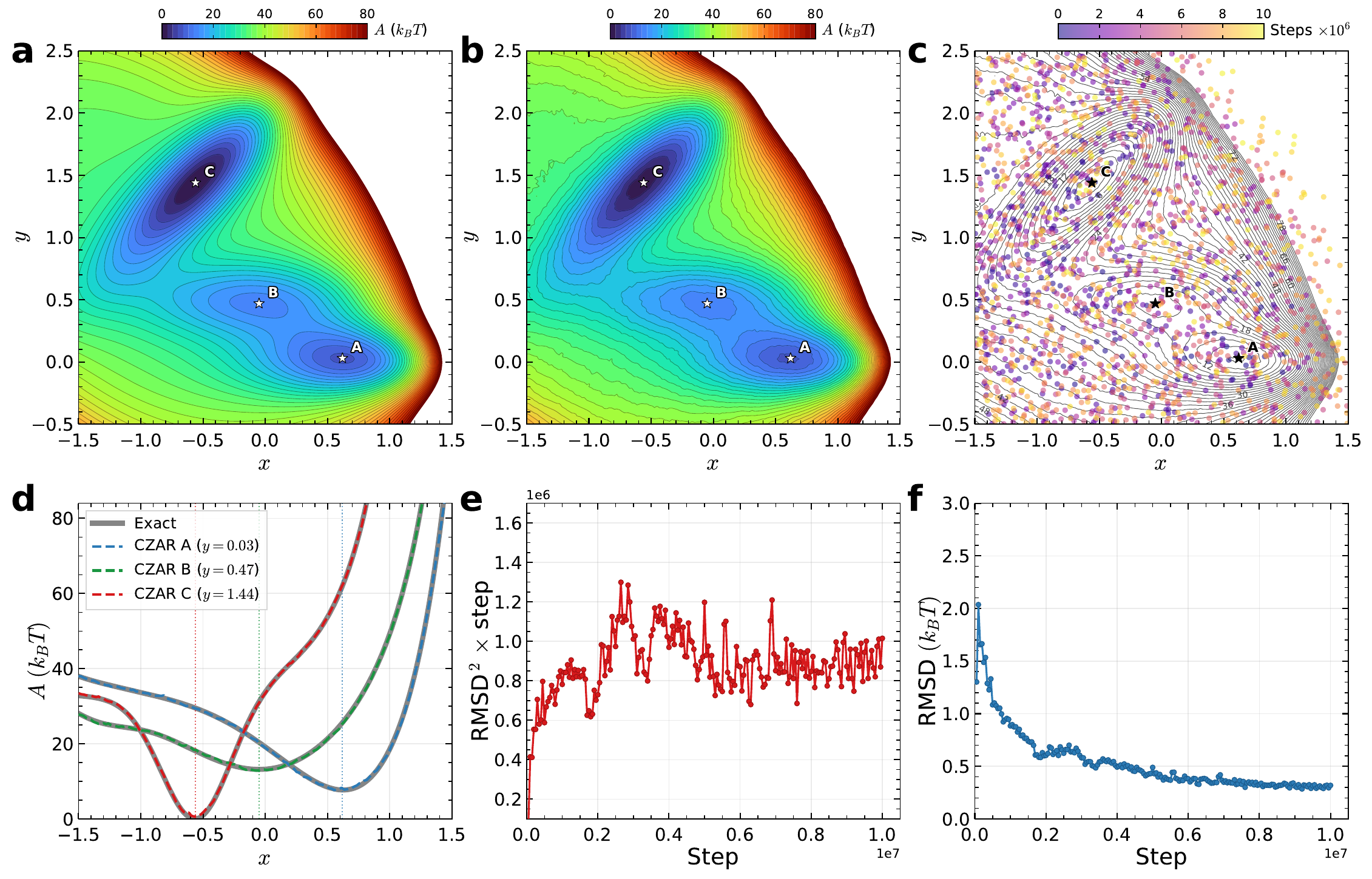}
\caption{\textbf{(a)} Analytical free energy $A(z)$ of the scaled M\"uller--Brown potential. \textbf{(b)} CZAR recovered free energy from FK-eABF after 10M steps. \textbf{(c)} Exploration trajectory (stride of 5000 steps) of the fictitious variable $\boldsymbol{\lambda}$ \textbf{(d)} 1D cross-sections through each basin minimum at a fixed $y$. Solid lines show the exact analytical profile, dashed colored lines show the CZAR estimate. \textbf{(e)} Product of $\rm{RMSD}^2$ and the step number, plotted against simulation time. \textbf{(f)} RMSD of the CZAR FEL relative to the analytical Muller--Brown surface, evaluated over all grid points within $80\,k_{\rm B}T$ of the global minimum.}
\label{fig:converged_fel}
\end{figure*}

The time evolution of the kernel population and reconstructed free energy is presented in \figref{fig:kernel_evolution}. At early simulation times ($10^5$ steps, \figref{fig:kernel_evolution}a), the CZAR-reconstructed FEL already captures the neighborhoods of all basins, and the local contour structure is qualitatively correct. The corresponding kernel population (\figref{fig:kernel_evolution}b) consists of a small number of wide kernels concentrated near basin C, reflecting the initially large Silverman bandwidth. The signed force components stored in these kernels (\figref{fig:kernel_evolution}c) already display a clear directional organization, with opposing signs on either side of the minimum. By $5\times10^5$ steps (\figref{fig:kernel_evolution}d--f), all three basins have been sampled, the kernel population has grown to 7263 compressed kernels with markedly narrower bandwidths, and the CZAR FEL correctly resolves all three basins. The gradient map now shows a coherent spatial structure, with well-defined sign changes along the $x$ direction that faithfully trace the underlying free-energy gradient. At $10^6$ steps (\figref{fig:kernel_evolution}g--i), the kernel population becomes denser (13,252 kernels) and the Silverman bandwidth has contracted toward $\sigma_{\min}$, yielding a nearly complete and quantitatively resolved FEL spanning all basins and barrier regions.

\begin{figure*}[!ht]
\centering
\includegraphics[width=\textwidth]{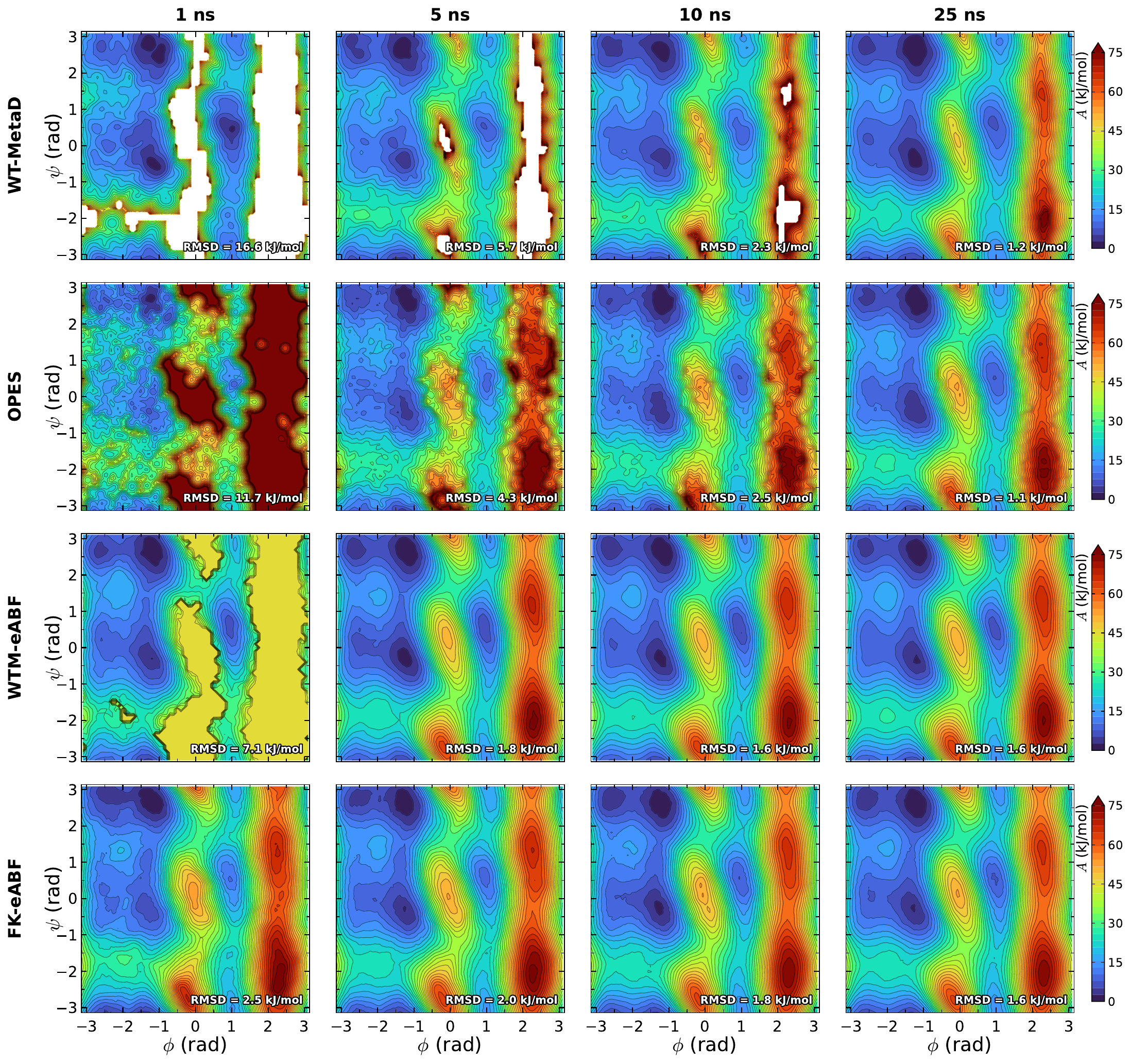}
\caption{Convergence of the FEL for $N$-acetyl-$N^\prime$-methylalanylamide in explicit water. Each row corresponds to a different enhanced sampling method: WT-MetaD, OPES, WTM-eABF, and FK-eABF. Columns show snapshots at 1, 5, 10, and 25 ns of simulation time. RMSD values (lower right of each panel, in kJ/mol) are computed against a converged, averaged, consensus FEL.}
\label{fig:adp}
\end{figure*}

\figref{fig:converged_fel} provides a direct comparison between the analytical M\"uller--Brown surface (\figref{fig:converged_fel}a) and the FK-eABF result after $10^7$ steps (\figref{fig:converged_fel}b). All three minima are correctly located, the contour spacing is consistent throughout the low-energy region, and the barrier ridges between basins are accurately reproduced. The trajectory of $\boldsymbol{\lambda}$, sampled every 5000 steps and overlaid on the CZAR contour lines (\figref{fig:converged_fel}c), reveals an accumulation near basin C followed by ergodic exploration of the full landscape. One-dimensional cross-sections through each basin minimum (\figref{fig:converged_fel}d) confirm quantitative agreement between the CZAR estimate (dashed colored lines) and the analytical profile (solid gray lines). Basin A ($y=0.03$) and basin B ($y=0.47$) are reproduced essentially exactly across the full cross-section, including the flanking barrier regions. Basin C ($y=1.44$), which corresponds to the initial basin, shows slightly larger, yet still sub-$k_{\rm B}T$ deviations at the minimum while preserving the correct well depth and curvature.

Figure~S1 further shows that the $\mathbf{z}$-kernel density, the associated density-gradient correction, and the Nadaraya--Watson mean-gradient field evaluated on the $\boldsymbol{\lambda}$-kernels remain spatially smooth and free of spurious discontinuities at early, intermediate, and converged stages of sampling. This comparison is not intended to reproduce the physical equilibrium density itself, but rather to demonstrate that the biased $z$-kernel density estimate entering the second CZAR term remains numerically stable as the kernel population grows. In addition, the $\mathbf{z}$-kernel population naturally extends beyond the walled $\boldsymbol{\lambda}$ domain by approximately $\sigma_{\min}$ because $\mathbf{z}$ fluctuates thermally around $\boldsymbol{\lambda}$, allowing the NW estimator to be evaluated across the entire reported CV range (Figure~S2).

The convergence behavior of the CZAR free-energy estimate is quantitatively assessed in \figref{fig:converged_fel}e. We monitor the product $\mathrm{RMSD}^2 \times t$,\cite{branduardi2012metadynamics,pfaendtner2015efficient} which plateaus when the error decays as $\mathrm{RMSD} \propto 1/\sqrt{t}$, the expected scaling for an unbiased estimator accumulating statistically independent mean-force samples. In FK-eABF, this plateau is reached after approximately $2\times10^6$ steps (\figref{fig:converged_fel}e), indicating that the estimator has entered the asymptotic convergence regime within the first quarter of the trajectory. The raw RMSD (\figref{fig:converged_fel}f) decreases monotonically from approximately 2.0 $k_{\rm B}T$ at $10^5$ steps to 0.7 $k_{\rm B}T$ at $5\times10^6$ steps, falling below $k_{\rm B}T$ well before mid-simulation. No systematic deviation from the $1/\sqrt{t}$ scaling is observed at late times.

\begin{figure*}[!ht]
\centering
\includegraphics[width=\textwidth]{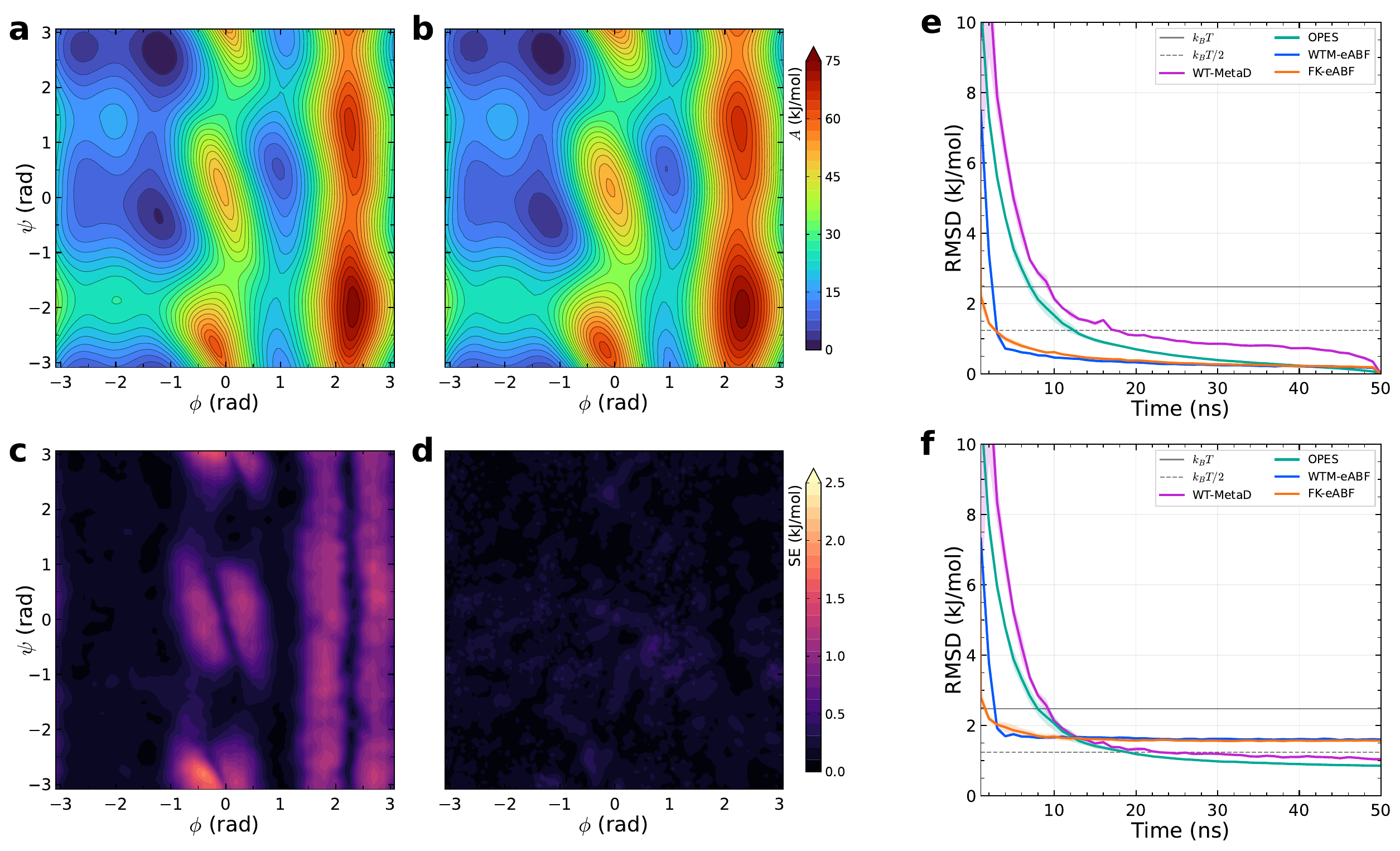}
\caption{Accuracy and convergence of the FEL for $N$-acetyl-$N^\prime$-methylalanylamide in explicit water. (a) Consensus FEL obtained by  averaging across all methods and replicas at 50 ns (three replicas for each of the four methods). (b) FK-eABF FEL averaged over three independent replicas at 50 ns. (c) Standard error of the consensus FEL from bootstrap resampling. (d) Standard error of the FK-eABF FEL across three replicas. (e) RMSD of each method relative to its own converged PMF (mean $\pm\rm{SE}$ over three replicas). (f) RMSD of each method relative to the cross-method consensus FEL.}
\label{fig:adp_conv}
\end{figure*}

\paragraph{\textbf{$N$-acetyl-$N^\prime$-methylalanylamide}.} To assess FK-eABF against established enhanced-sampling methods on a well-known molecular system, we benchmark the recovery of the Ramachandran FEL for NANMA, solvated in explicit TIP3P water\cite{Jorgensen1983}. Four methods are compared: WT-MetaD, OPES, WTM-eABF, and FK-eABF. All simulations use identical system preparation, settings, and force field (AMBER ff14SB\cite{maier2015ff14sb}). Three independent replicas of 50 ns each are performed for each of the four methods, and the computational overhead of all methods is evaluated on identical hardware and reported in Table S1. FK-eABF incurs negligible overhead despite the nearly 200-fold higher kernel deposition frequency relative to OPES. Full simulation details are  provided in the Supporting Information. 

\figref{fig:adp} shows the recovered FEL at 1, 5, 10, and 25 ns for a representative replica of each method. RMSD values are computed against a cross-method consensus FEL. The consensus FEL was constructed by converting each replica's  final-time FEL to a normalized probability distribution,  bootstrap-resampling across all twelve replicas (three per method,  four methods), and transforming back via $-k_BT \ln \langle p \rangle$.  This probability-space averaging avoids anchoring to any single reference state, and distributes uncertainty according  to inter-method variability across the full CV range rather than artificially concentrating it (details of both the bootstrapping procedure and convergence metrics are provided in the Supporting Information).

At 1.0 ns, the methods exhibit markedly different coverage and accuracy. WT-MetaD and WTM-eABF retain large unsampled regions where the free energy is undefined. OPES achieves greater coverage, but displays substantial errors in the barrier regions (RMSD of 11.7 kJ/mol). FK-eABF ($\gamma=10$) already yields a qualitatively correct FEL with full domain coverage and an RMSD of 2.5 kJ/mol, i.e., approximately $k_{\rm B}T$, demonstrating that the density-based exploration force rapidly drives ${\lambda}$ across all basins while the CZAR estimator extracts a quantitative FEL from early-time data. By 25 ns, all four methods converge to 1.1--1.6 kJ/mol RMSD against the consensus, confirming that they recover the same underlying surface given sufficient simulation time. The key distinction lies in the convergence rate: FK-eABF reaches approximately $k_{\rm B}T$ accuracy after 1.0 ns, a timescale at which WT-MetaD and WTM-eABF have not yet fully explored configurational space, and OPES remains unconverged in barrier regions.
\begin{figure*}[!ht]
\centering
\includegraphics[width=\textwidth]{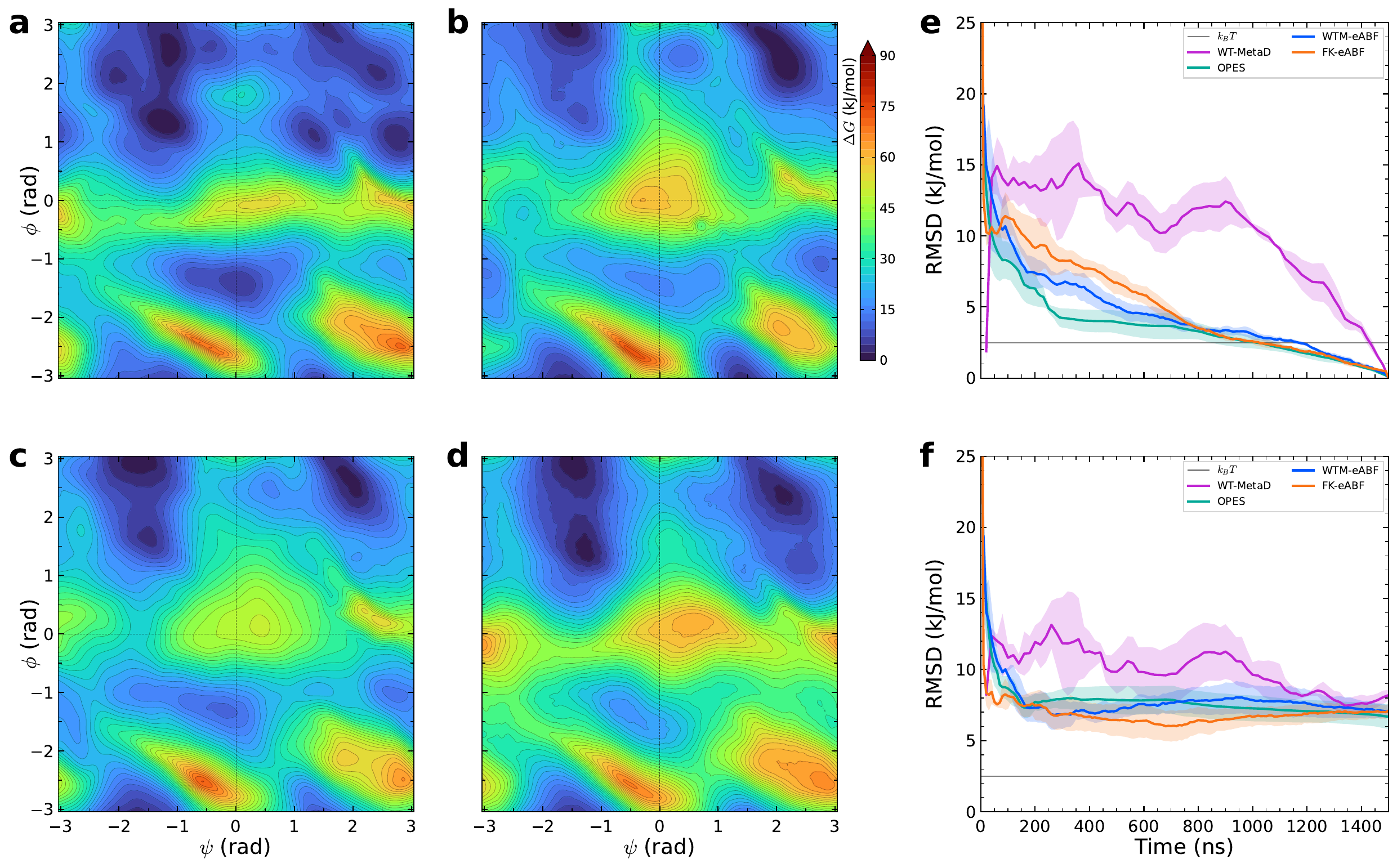}
\caption{FEL of the DFG-in/out transition in apo Abl1 kinase for (a) WT-MetaD, (b) OPES, (c) WTM-eABF, and (d) FK-eABF. (e) Self-convergence RMSD of the FEL for each method relative to its own final estimate, averaged over three replicas with shaded bands denoting the standard error. (f) Convergence to the cross-method consensus FEL.}
\label{fig:abl_pmfs}
\end{figure*}

A more detailed convergence analysis is presented in \figref{fig:adp_conv}. Panels (a) and (b) show the consensus FEL (defined as the average over all 12 simulations at 50 ns) and the FK-eABF FEL averaged over three replicas at 50 ns. Panels (c) and (d) of \figref{fig:adp_conv} show the uncertainty estimates over the respective FELs. The consensus SE (\figref{fig:adp_conv}c), computed by bootstrap resampling across all methods and replicas, ranges from 1.0 to 2.0~kJ/mol in the high-energy regions, and is below 0.5~kJ/mol in the metastable basins. The FK-eABF inter-replica SE (\figref{fig:adp_conv}d) is consistently low across the entire CV space, generally below 0.5~kJ/mol, indicating strong reproducibility across independent trajectories.

Panels (e) and (f) of \figref{fig:adp_conv} track the RMSD as a function of simulation time, averaged over three replicas with shaded bands denoting the standard error. Panel (e) shows self-convergence, where each method is compared against its own final PMF, isolating the intrinsic convergence rate of the estimator independent of cross-method differences. Panel (f) shows convergence to the cross-method consensus FEL. In both panels, FK-eABF (orange) reaches $k_{\rm B}T$-level accuracy within the first nanosecond and then decreases monotonically, reaching approximately 1.6 kJ/mol by 25 ns and remaining stable up to 50 ns. WTM-eABF (blue) follows a similar trend but with a higher initial RMSD due to the time required for histogram bins to accumulate sufficient counts across the full domain. However, once WTM-eABF achieves sufficient bin counts for a reliable force estimate, its RMSD decreases more steeply, indicating that the histogram-based estimator converges rapidly once adequate sampling is achieved locally. WT-MetaD (purple) converges steadily but more slowly, consistent with the gradual filling of free-energy basins by accumulated Gaussian bias. OPES (teal) exhibits convergence slightly faster than WT-MetaD, but starts from a significantly higher initial error, in excess of 10.0 kJ/mol, and requires approximately 9.0 ns to reach sub-$k_{\rm B}T$ accuracy. 

\paragraph{Abl1 kinase DFG-in/out transition}
The DFG (Asp–Phe–Gly)-in/out conformational flip of the Abl1 kinase domain provides a stringent validation case for FK-eABF, featuring a rugged FEL, with realistic barriers, and transitions coupled to slow orthogonal degrees of freedom. Shan~et~al.\cite{shan_conserved_2009} observed the DFG flip in long-timescale unbiased MD simulations and suggested that the DFG-in and DFG-out states are near isoenergetic at physiological pH. Oruganti~et~al.\cite{oruganti_allosteric_2022} later reported $\Delta G_{\text{in} \to \text{out}} \approx 0$~kJ/mol for the apo kinase using WT-MetaD along the same dihedral CVs used here. This established thermodynamic balance makes Abl1 an ideal system for testing whether FK-eABF can recover quantitatively accurate free-energy differences on a challenging two-dimensional landscape.

\begin{figure*}[!hb]
\centering
\includegraphics[width=\textwidth]{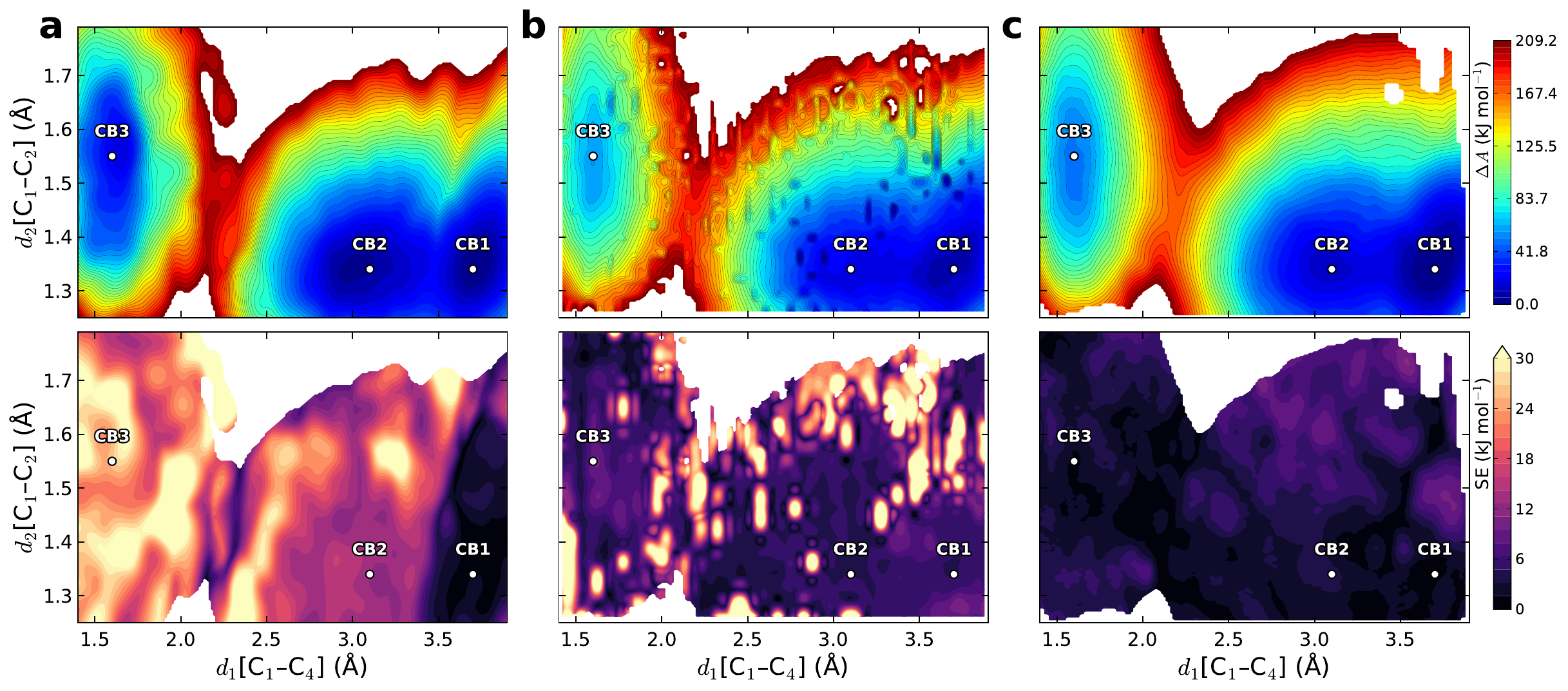}
\caption{Free-energy landscape and barrier convergence for the ring closure of 1,3-butadiene. Each column corresponds to one enhanced-sampling method: (a) OPES, (b) WTM-eABF, and (c) FK-eABF. The top row are the three-replica averaged FELs, with the corresponding bootstrap standard error shown below. White regions indicate energies exceeding the color-bar ceiling.}
\label{fig:btd_pmfs}
\end{figure*}
Three independent replicas of 1500 ns each were performed for each of the four methods on the 93,437-atom system, yielding a total of 18.0~$\mu$s of aggregate sampling under identical conditions (details in the Supporting Information). \figref{fig:abl_pmfs}a--d show the averaged FEL for each method. All four recover the same global topography: A DFG-out basin near $(\psi,\,\phi) \approx (-2.1,\,-2.4)$~rad, a DFG-in basin near $(-1.2,\,1.4)$~rad, and barrier regions located consistently across methods, although WT-MetaD (\figref{fig:abl_pmfs}a) exhibits residual noise indicative of slower convergence. The cross-method consensus and bootstrap standard error are reported in Figure S4.

Convergence is quantified in \figref{fig:abl_pmfs}e--f. FK-eABF reaches sub-$k_{\rm B}T$ self-convergence within ${\sim}$1000~ns and maintains low inter-replica variance throughout the trajectory. FK-eABF, OPES, and WTM-eABF all plateau near 7.0~kJ/mol RMSD relative to the consensus (\figref{fig:abl_pmfs}f). To quantify the thermodynamic balance between states, we integrated the Boltzmann-weighted probability over DFG-in and DFG-out basins defined following Oruganti~et~al.\cite{oruganti_allosteric_2022} The resulting free-energy differences are consistent with a near-isoenergetic equilibrium. The consensus FEL (Figure S4e) yields $\Delta G_{\text{in} \to \text{out}} = 1.0$~kJ/mol, in agreement with prior investigations. FK-eABF gives $\Delta G_{\text{in} \to \text{out}} = 0.3$~kJ/mol, well within $k_{\rm B}T$, while the values across all four methods span 0.3--8.2~kJ/mol, reflecting residual variability in sampling convergence across methods.

\paragraph{Electrocyclic Ring Closing of 1,3-Butadiene}

To evaluate the ability of FK-eABF in regimes where each MD step is computationally demanding and sampling efficiency is critical, we study the electrocyclic ring closure of 1,3-butadiene to cyclobutene using \textit{ab initio} MD (AIMD) simulations. In contrast to the Abl1 kinase application, where long aggregate simulation times are accessible but convergence is slowed by a rugged biomolecular landscape, the butadiene reaction probes whether FK-eABF can recover the correct free-energy topology under extreme sampling constraints of AIMD, where the total trajectory budget is limited to a few tens of picoseconds rather than microseconds.

Following the earlier study by Awasthi et al.\cite{awasthi2017exploring} we chose the distance between terminal carbon atoms (d$_1[\mathrm{C}_1 - \mathrm{C}_4]$) and the distance between the internal double-bonded carbons (d$_2[\mathrm{C}_2-\mathrm{C}_3]$) as CVs to model this reaction. 
The FEL obtained using these CVs identifies three distinct minima corresponding to \textit{trans}-1,3-butadiene (CB1), \textit{cis}-1,3-butadiene (CB2), and cyclobutene (CB3) as shown in \figref{fig:systems}. Since standard WT-MetaD was shown to be ineffective for this system, with free energy barriers failing to converge even after 1000~ps\cite{awasthi2016sampling,awasthi2017exploring}, we, therefore, restricted the comparison to OPES, WTM-eABF, and FK-eABF.

Three independent replicas of 30\,ps were performed for each method (simulation settings are provided in the Supporting Information). The replica-averaged FEL (top row) and bootstrap standard error (bottom row) for OPES (a), WTM-eABF (b), and FK-eABF (c) are shown in \figref{fig:btd_pmfs}. All three methods identify the expected basins~\cite{awasthi2017exploring} (CB1, CB2, and CB3) and the qualitative topology of the landscape is consistent across methods. However, under this stringent time constraint, the FELs differ markedly in smoothness, uncertainty, and convergence behavior (Figure S5). The FEL obtained using OPES (\figref{fig:btd_pmfs}a) displays pronounced noise, particularly in the barrier region, between CB2$\to$CB3. The standard error map also reveals substantial uncertainties that persist across the CV space. WTM-eABF (\figref{fig:btd_pmfs}b) shows considerably improved sampling relative to OPES, and all three basins are clearly visible in the free energy profile. However, the FEL exhibits noticeable artifacts and elevated standard errors, especially along d$_1[\mathrm{C}_1 - \mathrm{C}_4]$ CV near CB3. In contrast, FK-eABF (\figref{fig:btd_pmfs}c) recovers a smooth and well-resolved FEL with distinct minima and high energy regions. The standard error remains uniformly low across the CV space, indicating high reproducibility across independent replicas despite the short 30-ps timescale. Further, free-energy barriers stabilize in as little as 15 ps (Figure S5). These results demonstrate that FK-eABF achieves substantially faster convergence in regimes where OPES and WTM-eABF remain under-sampled.

\section{Discussion}

\subsection{Architecture}

The defining architectural property of FK-eABF is the replacement of fixed histogram bins with a compressed population of adaptive kernels that store running mean force estimates. By estimating and canceling the mean force, ABF flattens the effective landscape seen by $\boldsymbol{\lambda}$ and promotes broad exploration along the CV, allowing force samples to accumulate in regions that unbiased dynamics would visit only rarely.\cite{chipot2007free} Force samples are particularly well suited to this strategy because each sample contributes local gradient information immediately, rather than requiring a globally converged density estimate to stabilize over the full domain. The NW regression then spreads the information for each sample across a region set by the kernel bandwidth, yielding a smooth gradient field from the earliest stages of sampling, rather than waiting for histogram bins to become adequately populated.

Comer et al.\ proposed exactly this kind of kernel-based adaptation as an avenue for improving convergence in ABF, suggesting that the mean force in a given bin should depend not only on samples accrued in that bin but also on samples in neighboring bins.\cite{comer2015adaptive} They further observed that the best accuracy in ABF-derived free energies is obtained not when the free energy is perfectly flat, but when the statistical error of the mean force is approximately uniform along the transition coordinate. This perspective has more recently been formalized in so-called mean-force integration approaches that use kernel-weighted mean-force estimates and adopt the local standard error of the mean force as a primary convergence diagnostic.\cite{10.1063/1.5123498,bjola2024estimating} The NW regression in FK-eABF---formalized as a kernel-smoothed mean-force estimator for the ABF setting by Leli\`evre, Rousset, and Stoltz\cite{stoltz2010free} and rigorously analyzed for convergence by Jourdain, Leli\`evre, and Roux\cite{jourdain2010existence}---provides a concrete connection between these two ideas. First, by borrowing information across neighboring regions of CV space, it implements the kernel-coupled adaptation envisioned by Comer et al. Second, by smoothing the local mean-force estimate in sparsely sampled regions, it reduces estimator variance, thereby promoting a more uniform distribution of statistical error along the CV, and improving the accuracy of the recovered free energy.
 
A second consequence of the kernel representation is that the cancellation force and the optional exploration drive can be constructed from the same underlying data structure rather than from separate ones. FK-eABF inherits from eABF the property that the choice of bias on $\boldsymbol{\lambda}$ does not affect the asymptotic exactness of CZAR\cite{lesage2017,fu2019}, and like WTM-eABF it uses this freedom to augment the ABF cancellation force with an additional drive that accelerates early-time exploration. The two methods differ in how that drive is realized. WTM-eABF deposits a separate MetaD potential whose hills accumulate in an independent data structure on an independent schedule from the ABF accumulator. FK-eABF instead derives both contributions from a single kernel population: the cancellation force from the NW numerator-over-denominator ratio, and the exploration force from the gradient of the same kernel-density estimate that appears in the denominator. As a result, the two contributions to the bias on $\boldsymbol{\lambda}$ are intrinsically coupled: they share the same kernel population and are updated synchronously at each grid rebuild. An analogous density-driven idea underlies OPES, which constructs its bias from a kernel-density estimate to target a chosen distribution, typically the well-tempered one.\cite{invernizzi_rethinking_2020} FK-eABF differs in the role assigned to the density estimate: it drives only the auxiliary exploration force on $\boldsymbol{\lambda}$, while the free-energy estimator is governed independently by CZAR applied to the $\mathbf{z}$-kernel population. This separation preserves CZAR's asymptotic exactness regardless of the shape or convergence state of the exploration drive.

\subsection{Performance across timescales}

The M\"uller--Brown benchmark establishes the correctness of the kernel CZAR formalism against an exact analytical reference (\figref{fig:converged_fel}). On NANMA in explicit water, the agreement at long times with WT-MetaD, OPES, and WTM-eABF demonstrates that the kernel representation does not introduce a systematic offset relative to established approaches, while the clear early-time separation witnessed in \figref{fig:adp_conv} is consistent with the architectural properties described above (Figure~S3).

A more demanding test is whether this early-time advantage persists on a rugged biomolecular landscape with slow orthogonal relaxation. For the Abl1 kinase DFG-in/out transition, all four methods recover a consistent global topography (\figref{fig:abl_pmfs}; Figure~S4), including the locations of the DFG-in and DFG-out basins and the intervening barriers. FK-eABF also reproduces the near-isoenergetic balance between DFG-in and DFG-out reported previously for this system. At the same time, the spread of $\Delta G_{\mathrm{in}\to\mathrm{out}}$ values across methods (0.3--8.2~kJ/mol) indicates that the landscape remains sensitive to estimator choice and residual sampling error, even after microsecond-scale sampling. The main result here is therefore not that FK-eABF yields a uniquely correct surface, but that it converges to a solution consistent with both the cross-method consensus and prior literature, while maintaining low inter-replica variability. 

The strongest practical advantage emerges in the AIMD regime, where the total sampling budget is limited to tens of picoseconds. For the electrocyclic ring closure of 1,3-butadiene, FK-eABF recovers a smooth and well-resolved free-energy landscape with low bootstrap uncertainty within 30~ps, whereas OPES and WTM-eABF remain visibly noisier under the same sampling budget, and exhibit less stable barrier estimates (Figure~S5). The ability to construct a reliable mean-force estimate without fully populated histogram bins is particularly valuable here, and these results suggest that FK-eABF is especially well suited to problems characterized by high per-step computational cost and limited accessible trajectory length.

\subsection{Practical considerations}
Relative to histogram-based eABF, FK-eABF introduces only a small number of additional parameters, and these remain directly physically interpretable. The most important are the initial bandwidth $\sigma_0$ and the minimum bandwidth $\sigma_{\min}$. In practice, a useful starting point is to choose $\sigma_0$ on the order of the histogram bin width one would have used for a conventional eABF calculation on the same CV, and to set $\sigma_{\min}$ to roughly half that value. The bandwidth floor is essential for stable and efficient behavior; without it, continued contraction of the kernel population increases memory use and neighbor-search cost without yielding meaningful gain in resolution beyond the noise scale of the sampled mean-force field.

The exploration parameter $\gamma$ controls the strength of the density-based repulsion that drives early-stage coverage of the CV domain. Its influence naturally diminishes as sampling becomes more uniform and the ABF component converges, so it requires less delicate tuning than the well-tempered bias factor in metadynamics-based methods. For NANMA, values of $\gamma \le 10$ were sufficient to obtain robust convergence (Figure~S6), and across all systems studied here effective values remained in a moderate range ($\gamma=8~\rm{to}~12$). The method was likewise robust across the tested PACE range (Figures~S7--S9), suggesting that more frequent data acquisition can improve convergence under stringent sampling constraints, like in AIMD, but need not be fine-tuned with high precision in general applications.

Just like with other kernel-based estimators, boundary behavior depends on a sensible choice of CV domain. As walls are applied only to $\boldsymbol{\lambda}$ and not to $\mathbf{z}$ directly, the $\mathbf{z}$-kernel population naturally extends slightly beyond the nominal CV domain owing to harmonic coupling (Figure~S2), so points inside the reported CV range retain approximately symmetric kernel support. In the benchmarks studied here, this proved sufficient to prevent boundary artifacts, such as kernel accumulation, or systematic drift, including for the distance-based butadiene coordinates. This should, nonetheless, be viewed as a practical safeguard rather than a formal guarantee. Poorly chosen CV bounds can still degrade kernel-based estimators, and domain selection requires the same level of care as in any other free-energy method.

\subsection{Conclusion} 
Three conclusions can be drawn from the present work. First, replacing histogram accumulators with a compressed kernel population carrying running-mean forces yields smooth, threshold-free mean-force estimates from the earliest stages of sampling, while preserving long-time accuracy comparable to established enhanced-sampling approaches. Second, the resulting gain is most pronounced when the sampling budget is severely constrained, like in AIMD, where the method can produce a quantitatively useful landscape before histogram-based or hill-based approaches have fully stabilized. Third, the same algorithmic framework operates robustly across more than four orders of magnitude in simulation time, from 30~ps of AIMD for butadiene cyclization to 1.5~$\mu$s of classical MD for the Abl1 DFG transition, with only modest adjustment of a small set of physically interpretable parameters.

More broadly, FK-eABF should be viewed not simply as a kernelized implementation of eABF, but as a reformulation of the underlying eABF architecture, in which kernel representations are used for both mean-force estimation and auxiliary exploration. This reformulation suggests a natural extension of the broader $x$ABF family, in which an ABF-like mean-force term along an extended coordinate is paired with a coordinate-appropriate recovery procedure, e.g., CZAR for geometrical variables, thermodynamic integration for alchemical ones, and integration- or reweighting-based schemes for temperature-like variables.\cite{lagardere2024lambda,zhou2026,fu2020} Methods in this family differ both in the choice of extended coordinate and in how early-time exploration is added on top of ABF cancellation: WTM-$x$ABF augments it with a WT-MetaD bias\cite{fu2020,zhou2026} whereas OPES-based eABF and $\lambda$ABF variants replace this deposited bias layer with a density-based exploration term reconstructed from an evolving kernel-density estimate. \cite{hulm2025combining,ansari2025lambda} FK-eABF extends this progression one step further by replacing the ABF accumulator itself with a force-kernel representation from which the exploration force is also reconstructed. In this view, an FK $x$ABF formulation could naturally supplant the separate MetaD component, rather than merely supplement it, using the kernel-derived exploration force as the sole auxiliary drive. A more aggressive FK-WTM-$x$ABF hybrid would still be possible if one wished to retain both exploration layers, but the two mechanisms would then be partially redundant and would require careful control of parameter coupling. The simpler MetaD-free variant therefore appears especially attractive for continuous alchemical and hybrid protocols in which accurate and local mean-force estimation remains central.

Taken together, these results identify FK-eABF as a compelling addition to the eABF family for low-dimensional CV problems in which early-time convergence is critical. Whether the same advantages persist in more strongly nonlocal or higher-dimensional CV spaces, and whether the FK-$x$ABF extension delivers comparable benefits for alchemical and temperature-like coordinates, remain open questions and important directions for future investigation. A natural next step is a multi-walker extension, which the associative kernel compression algebra accommodates straightforwardly and which is particularly attractive in the AIMD regime.

\section{Acknowledgments}    
C.K. acknowledges Rui Sun, Kazuumi Fujioka and Tony Leli\`evre for the insightful discussions regarding this work. C.C. acknowledges the European Research Council (project 101097272 ``MilliInMicro'')

\section{Supporting Information} 
Additional implementation details for FK-eABF are provided, including the kernel architecture, pairwise variance update procedure, KDE normalization factor derivation, recursive merge cascade, neighbor-list acceleration, and force-clamping bounds, along with computational overhead benchmarks. The construction of the cross-method consensus FEL is described in full, including the probability-space bootstrap procedure and the self convergence and consensus RMSD metrics used throughout. Sensitivity analyses for the exploration bias factor $\gamma$ and kernel deposition stride PACE, the per-method and consensus free-energy landscapes for Abl1 kinase, barrier-height convergence for 1,3-butadiene, and complete simulation parameters for all systems and methods are also provided. 

\section{Data Availability}
Input files, simulation parameters, and analysis scripts for all systems are available at \url{https://github.com/aditya1707/ForceKernel-eABF} (to be made public upon acceptance). The FK-eABF implementation is distributed as a standalone PLUMED module in the same repository.

\section{Conflicts of Interest}
The authors declare no competing financial interest.

\clearpage
\bibliography{references}

@Article{Torrie1977a,
  author  = {Torrie, G. M. and Valleau, J. P.},
  title   = {Nonphysical sampling distributions in Monte Carlo free energy estimation: Umbrella sampling},
  pages   = {187-199},
  volume  = {23},
  journal = {J. Comput. Phys.},
  year    = {1977},
}

@ARTICLE{Darve2001,
  author = {Darve, E. and Pohorille, A.},
  title = {Calculating free energies using average force},
  journal = {J. Chem. Phys.},
  year = {2001},
  volume = {115},
  pages = {9169-9183}
}

@Article{Laio2002,
  author  = {Laio, A. and Parrinello, M.},
  title   = {Escaping free energy minima},
  pages   = {12562-12565},
  volume  = {99},
  comment = {Free energy},
  journal = {Proc. Natl. Acad. Sci. U.S.A.},
  year    = {2002},
}

@article{Jorgensen1983,
  author = {William L. Jorgensen and Jayaraman Chandrasekhar and Jeffry D. Madura and Roger W. Impey and Michael L. Klein},
  title = {Comparison of simple potential functions for simulating liquid water},
  journal = {J. Chem. Phys.},
  volume = {79},
  pages = {926--935},
  year = {1983}
}

@article{invernizzi_rethinking_2020,
	title = {Rethinking Metadynamics: From Bias Potentials to Probability Distributions},
	volume = {11},
	issn = {1948-7185, 1948-7185},
	url = {https://pubs.acs.org/doi/10.1021/acs.jpclett.0c00497},
	doi = {10.1021/acs.jpclett.0c00497},
	shorttitle = {Rethinking Metadynamics},
	pages = {2731--2736},
	number = {7},
	journal = {J. Phys. Chem. Lett.},
    year = 2020,
	author = {Invernizzi, Michele and Parrinello, Michele},
	urldate = {2020-09-30},
	date = {2020-04-02},
}

@article{oruganti_allosteric_2022,
  title={Allosteric enhancement of the BCR-Abl1 kinase inhibition activity of nilotinib by cobinding of asciminib},
  author={Oruganti, Baswanth and Lindahl, Erik and Yang, Jingmei and Amiri, Wahid and Rahimullah, Rezwan and Friedman, Ran},
  journal={J. Biol. Chem.},
  volume={298},
  number={8},
  year={2022},
  publisher={Elsevier}
}

@article{shan_conserved_2009,
  title={A conserved protonation-dependent switch controls drug binding in the Abl kinase},
  author={Shan, Yibing and Seeliger, Markus A and Eastwood, Michael P and Frank, Filipp and Xu, Huafeng and Jensen, Morten {\O} and Dror, Ron O and Kuriyan, John and Shaw, David E},
  journal={Proc. Natl. Acad. Sci. U.S.A.},
  volume={106},
  number={1},
  pages={139--144},
  year={2009},
  publisher={National Academy of Sciences}
}

@article{mora2013,
  title={Mechanistic picture for conformational transition of a membrane transporter at atomic resolution},
  author={Moradi, Mahmoud and Tajkhorshid, Emad},
  journal={Proc. Natl. Acad. Sci. U.S.A.},
  volume={110},
  number={47},
  pages={18916--18921},
  year={2013},
  publisher={National Academy of Sciences}
}

@article{barducci2008,
  title = {Well-Tempered Metadynamics: A Smoothly Converging and Tunable Free-Energy Method},
  author = {Barducci, Alessandro and Bussi, Giovanni and Parrinello, Michele},
  year = 2008,
  journal = {Phys. Rev. Lett.},
  volume = {100},
  number = {2},
  pages = {020603},
  doi = {10.1103/PhysRevLett.100.020603}
}

@article{fu2016,
  title = {Extended Adaptive Biasing Force Algorithm. An On-the-Fly Implementation for Accurate Free-Energy Calculations},
  author = {Fu, Haohao and Shao, Xueguang and Chipot, Christophe and Cai, Wensheng},
  year = 2016,
  journal = {J. Chem. Theory Comput.},
  volume = {12},
  number = {8},
  pages = {3506--3513},
  doi = {10.1021/acs.jctc.6b00447}
}

@article{fu2019,
  title = {Taming Rugged Free Energy Landscapes Using an Average Force},
  author = {Fu, Haohao and Shao, Xueguang and Cai, Wensheng and Chipot, Christophe},
  year = 2019,
  journal = {Acc. Chem. Res.},
  volume = {52},
  number = {11},
  pages = {3254--3264},
  doi = {10.1021/acs.accounts.9b00473}
}

@article{fu2020,
  title = {Finding an Optimal Pathway on a Multidimensional Free-Energy Landscape},
  author = {Fu, Haohao and Chen, Haochuan and Wang, Xin'ao and Chai, Hao and Shao, Xueguang and Cai, Wensheng and Chipot, Christophe},
  year = 2020,
  journal = {J. Chem. Inf. Model.},
  volume = {60},
  number = {11},
  pages = {5366--5374},
  doi = {10.1021/acs.jcim.0c00279}
}

@article{lesage2017,
  title = {Smoothed Biasing Forces Yield Unbiased Free Energies with the Extended-System Adaptive Biasing Force Method},
  author = {Lesage, Adrien and Leli{\`e}vre, Tony and Stoltz, Gabriel and H{\'e}nin, J{\'e}r{\^o}me},
  year = 2017,
  journal = {J. Phys. Chem. B},
  volume = {121},
  number = {15},
  pages = {3676--3685},
  doi = {10.1021/acs.jpcb.6b10055}
}

@article{fu2021,
  title = {BFEE2: Automated, Streamlined, and Accurate Absolute Binding Free-Energy Calculations},
  author = {Fu, Haohao and Chen, Haochuan and Cai, Wensheng and Shao, Xueguang and Chipot, Christophe},
  year = 2021,
  journal = {J. Chem. Inf. Model.},
  volume = {61},
  number = {5},
  pages = {2116--2123},
  doi = {10.1021/acs.jcim.1c00269}
}

@article{gumbart2013standard,
  title={Standard binding free energies from computer simulations: What is the best strategy?},
  author={Gumbart, James C and Roux, Beno{\^\i}t and Chipot, Christophe},
  journal={J. Chem. Theory Comput.},
  volume={9},
  number={1},
  pages={794--802},
  year={2013},
  publisher={ACS Publications}
}

@article{tse2018link,
  title={Link between membrane composition and permeability to drugs},
  author={Tse, Chi Hang and Comer, Jeffrey and Wang, Yi and Chipot, Christophe},
  journal={J. Chem. Theory Comput.},
  volume={14},
  number={6},
  pages={2895--2909},
  year={2018},
  publisher={ACS Publications}
}

@article{zhou2026,
  title = {One for All, All for One: A Unified Framework for Free-Energy Calculations},
  author = {Zhou, Mengchen and Liu, Xuyang and Shao, Xueguang and Chipot, Christophe and Cai, Wensheng and Fu, Haohao},
  date = {2026-01-06},
  journal = {Acc. Chem. Res.},
  volume = {59},
  number = {1},
  pages = {90--102},
  year={2026},
  doi = {10.1021/acs.accounts.5c00666}
}

@book{silverman2018density,
  title={Density estimation for statistics and data analysis},
  author={Silverman, Bernard W},
  year={2018},
  publisher={Routledge}
}

@article{kang2024impact,
  title={Impact of the unstirred water layer on the permeation of small-molecule drugs},
  author={Kang, Christopher and Shoji, Alyson and Chipot, Christophe and Sun, Rui},
  journal={Journal of Chemical Information and Modeling},
  volume={64},
  number={3},
  pages={933--943},
  year={2024},
  publisher={ACS Publications}
}

@article{bussi2020using,
  title={Using metadynamics to explore complex free-energy landscapes},
  author={Bussi, Giovanni and Laio, Alessandro},
  journal={Nature Reviews Physics},
  volume={2},
  number={4},
  pages={200--212},
  year={2020},
  publisher={Nature Publishing Group UK London}
}

@article{dama2014well,
  title={Well-tempered metadynamics converges asymptotically},
  author={Dama, James F and Parrinello, Michele and Voth, Gregory A},
  journal={Physical review letters},
  volume={112},
  number={24},
  pages={240602},
  year={2014},
  publisher={APS}
}

@article{branduardi2012metadynamics,
  title={Metadynamics with adaptive Gaussians},
  author={Branduardi, Davide and Bussi, Giovanni and Parrinello, Michele},
  journal={Journal of chemical theory and computation},
  volume={8},
  number={7},
  pages={2247--2254},
  year={2012},
  publisher={ACS Publications}
}

@article{pfaendtner2015efficient,
  title={Efficient sampling of high-dimensional free-energy landscapes with parallel bias metadynamics},
  author={Pfaendtner, Jim and Bonomi, Massimiliano},
  journal={Journal of chemical theory and computation},
  volume={11},
  number={11},
  pages={5062--5067},
  year={2015},
  publisher={ACS Publications}
}

@article{hulm2025combining,
  title={Combining Fast Exploration with Accurate Reweighting in the OPES-eABF Hybrid Sampling Method},
  author={Hulm, Andreas and Schiller, Robert P and Ochsenfeld, Christian},
  journal={Journal of Chemical Theory and Computation},
  volume={21},
  number={13},
  pages={6434--6445},
  year={2025},
  publisher={ACS Publications}
}

@article{zhou2025one,
  title={One for All, All for One: A Unified Framework for Free-Energy Calculations},
  author={Zhou, Mengchen and Liu, Xuyang and Shao, Xueguang and Chipot, Christophe and Cai, Wensheng and Fu, Haohao},
  journal={Accounts of Chemical Research},
  volume={59},
  number={1},
  pages={90--102},
  year={2025},
  publisher={ACS Publications}
}

@article{watson1964smooth,
  title={Smooth regression analysis},
  author={Watson, Geoffrey S},
  journal={Sankhy{\=a}: The Indian Journal of Statistics, Series A},
  pages={359--372},
  year={1964},
  publisher={JSTOR}
}

@article{nadaraya1964estimating,
  title={On estimating regression},
  author={Nadaraya, Elizbar A},
  journal={Theory of Probability \& Its Applications},
  volume={9},
  number={1},
  pages={141--142},
  year={1964},
  publisher={SIAM}
}

@article{chan1983algorithms,
  title={Algorithms for computing the sample variance: Analysis and recommendations},
  author={Chan, Tony F and Golub, Gene H and LeVeque, Randall J},
  journal={The American Statistician},
  volume={37},
  number={3},
  pages={242--247},
  year={1983},
  publisher={Taylor \& Francis}
}

@article{tribello2014plumed,
  title={PLUMED 2: New feathers for an old bird},
  author={Tribello, Gareth A and Bonomi, Massimiliano and Branduardi, Davide and Camilloni, Carlo and Bussi, Giovanni},
  journal={Computer physics communications},
  volume={185},
  number={2},
  pages={604--613},
  year={2014},
  publisher={Elsevier}
}

@article{maier2015ff14sb,
  title={ff14SB: improving the accuracy of protein side chain and backbone parameters from ff99SB},
  author={Maier, James A and Martinez, Carmenza and Kasavajhala, Koushik and Wickstrom, Lauren and Hauser, Kevin E and Simmerling, Carlos},
  journal={Journal of chemical theory and computation},
  volume={11},
  number={8},
  pages={3696--3713},
  year={2015},
  publisher={ACS Publications}
}

@article{meng2015computational,
  title={Computational study of the “DFG-flip” conformational transition in c-Abl and c-Src tyrosine kinases},
  author={Meng, Yilin and Lin, Yen-lin and Roux, Beno{\^\i}t},
  journal={The Journal of Physical Chemistry B},
  volume={119},
  number={4},
  pages={1443--1456},
  year={2015},
  publisher={ACS Publications}
}

@article{awasthi2017exploring,
  title={Exploring high dimensional free energy landscapes: Temperature accelerated sliced sampling},
  author={Awasthi, Shalini and Nair, Nisanth N},
  journal={The Journal of Chemical Physics},
  volume={146},
  number={9},
  year={2017},
  publisher={AIP Publishing}
}

@article{awasthi2016sampling,
  title={Sampling free energy surfaces as slices by combining umbrella sampling and metadynamics},
  author={Awasthi, Shalini and Kapil, Venkat and Nair, Nisanth N},
  journal={Journal of Computational Chemistry},
  volume={37},
  number={16},
  pages={1413--1424},
  year={2016},
  publisher={Wiley Online Library}
}

@book{chipot2007free,
  title={Free energy calculations},
  author={Chipot, Christophe and Pohorille, Andrew},
  volume={86},
  year={2007},
  publisher={Springer}
}

@article{comer2015adaptive,
  title={The adaptive biasing force method: Everything you always wanted to know but were afraid to ask},
  author={Comer, Jeffrey and Gumbart, James C and H{\'e}nin, J{\'e}r{\^o}me and Leli{\`e}vre, Tony and Pohorille, Andrew and Chipot, Christophe},
  journal={The Journal of Physical Chemistry B},
  volume={119},
  number={3},
  pages={1129--1151},
  year={2015},
  publisher={ACS Publications}
}

@article{10.1063/1.5123498,
    author = {Marinova, Veselina and Salvalaglio, Matteo},
    title = {Time-independent free energies from metadynamics via mean force integration},
    journal = {The Journal of Chemical Physics},
    volume = {151},
    number = {16},
    pages = {164115},
    year = {2019},
    month = {10},
    issn = {0021-9606},
    doi = {10.1063/1.5123498},
    url = {https://doi.org/10.1063/1.5123498},
    eprint = {https://pubs.aip.org/aip/jcp/article-pdf/doi/10.1063/1.5123498/15566232/164115_1_online.pdf},
}

@article{bjola2024estimating,
  title={Estimating free-energy surfaces and their convergence from multiple, independent static and history-dependent biased molecular-dynamics simulations with mean force integration},
  author={Bjola, Antoniu and Salvalaglio, Matteo},
  journal={Journal of Chemical Theory and Computation},
  volume={20},
  number={13},
  pages={5418--5427},
  year={2024},
  publisher={ACS Publications}
}

@article{lagardere2024lambda,
  title={Lambda-abf: Simplified, portable, accurate, and cost-effective alchemical free-energy computation},
  author={Lagard{\`e}re, Louis and Maurin, Lise and Adjoua, Olivier and El Hage, Krystel and Monmarch{\'e}, Pierre and Piquemal, Jean-Philip and H{\'e}nin, J{\'e}r{\^o}me},
  journal={Journal of Chemical Theory and Computation},
  volume={20},
  number={11},
  pages={4481--4498},
  year={2024},
  publisher={ACS Publications}
}

@article{ansari2025lambda,
  title={Lambda-ABF-OPES: Faster convergence with high accuracy in alchemical free energy calculations},
  author={Ansari, Narjes and Jing, Zhifeng Francis and Gagelin, Antoine and H{\'e}din, Florent and Aviat, F{\'e}lix and H{\'e}nin, J{\'e}r{\^o}me and Piquemal, Jean-Philip and Lagard{\`e}re, Louis},
  journal={The Journal of Physical Chemistry Letters},
  volume={16},
  number={19},
  pages={4626--4634},
  year={2025},
  publisher={ACS Publications}
}

@article{lindorff2011fast,
  title={How fast-folding proteins fold},
  author={Lindorff-Larsen, Kresten and Piana, Stefano and Dror, Ron O and Shaw, David E},
  journal={Science},
  volume={334},
  number={6055},
  pages={517--520},
  year={2011},
  publisher={American Association for the Advancement of Science}
}

@article{zimmerman2021sars,
  title={SARS-CoV-2 simulations go exascale to capture spike opening and reveal cryptic pockets across the proteome},
  author={Zimmerman, Maxwell I and Bowman, Gregory},
  journal={Biophysical Journal},
  volume={120},
  number={3},
  pages={299a},
  year={2021},
  publisher={Elsevier}
}

@article{jourdain2010existence,
  title={Existence, uniqueness and convergence of a particle approximation for the AdaptiveBiasing Force process},
  author={Jourdain, Benjamin and Leli{\`e}vre, Tony and Roux, Rapha{\"e}l},
  journal={ESAIM: Mathematical Modelling and Numerical Analysis},
  volume={44},
  number={5},
  pages={831--865},
  year={2010},
  publisher={EDP Sciences}
}

@book{stoltz2010free,
  title={Free energy computations: A mathematical perspective},
  author={Stoltz, Gabriel and Rousset, Mathias and others},
  year={2010},
  publisher={World Scientific}
}

\end{document}